\newcounter{myctr}
\def\myitem{\refstepcounter{myctr}\bibfont\noindent\ifnum\themyctr>9\else\phantom{0}\fi\hangindent17pt\themyctr.\enskip}

\documentclass{ws-ijqi}
\newcommand{\ket}[1]{\left\vert#1\right\rangle}

\newcommand{\bra}[1]{\left\langle#1\right\vert}

\newcommand{\nbar}{\overline{n}}

	\newcommand{\tr}[1]{\textrm{Tr} \left[ {#1} \right]} % Faster traces

	\newcommand{\e}[1]{e^{ {#1}}} % Faster exponentials

\begin{document}

\markboth{L. Mazzola, G. De Chiara, and M. Paternostro}
{Detecting the work statistics through Ramsey-like interferometry}

%%%%%%%%%%%%%%%%%%%%% Publisher's Area please ignore %%%%%%%%%%%%%%
\catchline{}{}{}{}{}
%%%%%%%%%%%%%%%%%%%%%%%%%%%%%%%%%%%%%%%%%%%%%%%%%%%%%%%%%%%%%%%%%%%

\title{Detecting the work statistics through Ramsey-like interferometry}

\author{Laura Mazzola}

\address{Centre for Theoretical Atomic, Molecular and Optical Physics, School of Mathematics and Physics, Queen's University, Belfast BT7 1NN, United Kingdom\\
l.mazzola@qub.ac.uk}

\author{Gabriele De Chiara}

\address{Centre for Theoretical Atomic, Molecular and Optical Physics, School of Mathematics and Physics, Queen's University, Belfast BT7 1NN, United Kingdom\\
g.dechiara@qub.ac.uk}

\author{Mauro Paternostro}

\address{Centre for Theoretical Atomic, Molecular and Optical Physics, School of Mathematics and Physics, Queen's University, Belfast BT7 1NN, United Kingdom\\
m.paternostro@qub.ac.uk}

\maketitle

\begin{history}
\received{\today}
%\revised{??}
%\accepted{Day Month Year}
%\comby{(xxxxxxxxxx)}
\end{history}

\begin{abstract}
Out-of-equilibrium statistical mechanics is attracting considerable interest due to the recent advances in the control and manipulations of systems at the quantum level. Recently, an interferometric scheme for the detection of the characteristic function of the work distribution following a time-dependent process has been proposed [L. Mazzola {\it et al}, Phys. Rev. Lett. {\bf 110} 230602 (2013)]. There, it was demonstrated that the work statistics of a quantum system undergoing a process can be reconstructed by effectively mapping the characteristic function of work on the state of an ancillary qubit. Here, we expand that work in two important directions. We first apply the protocol to an interesting specific physical example consisting of a superconducting qubit dispersively coupled to the field of a microwave resonator, thus enlarging the class of situations for which our scheme would be key in the task highlighted above. We then account for the interaction of the system with an additional one (which might embody an environment), and generalise the protocol accordingly. %Thirdly, we contemplate the case where a harmonic oscillator is a practical ancilla to couple to the system and show the usefulness of the protocols also in this case and the technical changes that need to be made.
\end{abstract}

\keywords{Work statistics; Interferometry; Matter-light interaction.}

\section{Introduction}	%) A SECTION HEADING

The assessment of out-of-equilibrium statistics of quantum systems subjected to time-dependent processes is attracting an increasing degree of attention from the community interested in modern quantum physics\cite{CampisiRMP}. The Crooks and Jarzynski relations\cite{Tasaki,Crooks,Jarzynski}, which take into account fluctuations in non-equilibrium dynamics, connect thermodynamical properties at equilibrium to the non-equilibrium details of dynamics. The verification of their quantum mechanical counterparts has so far encountered substantial difficulties due to the practical difficulty to perform reliable projective measurements of instantaneous energy states\cite{CampisiRMP,Huber}, which are steps required in order to fully reconstruct the statistics of work.

In Refs.~\refcite{Dorner} and \refcite{Mazzola}, a radical change to the approach for the reconstruction of the work statistics has been proposed, inspired by phase-estimation protocols that are well-known in quantum information processing. The method, which relies on the use of a {\it clean} and controllable ancilla, suitably coupled to the system of interest, has very recently enabled the first experimental characterization of quantum fluctuation relations\cite{Batalhao}. Together with more recent schemes designed to address the quantum scenario\cite{Huber,Heyl,Pekola}, this has embodied a significant complement to past experimental successful verifications of out-of-equilibrium fluctuation relations in classical systems\cite{Collin,Liphardt,Saira,Toyabe,Douarche}.

%In Ref.~\refcite{Mazzola} we proposed an interferometric scheme for the detection of the statistics of work of a quantum system affected by an external driving. The key idea of that scheme is that by introducing an ancillary system suitably coupled to the system of interest, it is possible to map on the state of the ancilla and retrieve the characteristic function of the work probability distribution.
In this paper we extend the discussion presented in Ref.~\refcite{Mazzola} by emphasising the versatility of the proposed interferometric approach to the reconstruction of the characteristic function and apply it to the study of the statistics of work done by an external driving potential that changes the frequency of a harmonic oscillator. This problem is key in the current theoretical design of Otto cycles based on trapped-ion technology\cite{Abah}, and this physical situation is indeed encountered in a number of experimental scenarios, from cavity-quantum electrodynamics to its superconducting-circuit counterpart. %Such example complements those presented in Ref.~\refcite{Mazzola}, we the interferometric scheme was applied to the case of a harmonic oscillator subjected to an external driving that displaced it. 

The remainder of this paper is organised as followed. In Sec.~\ref{description} we give a brief review of the interferometric scheme at the core of our analysis. Sec.~\ref{Example} illustrates its application to the physical situation depicted above. In Sec.~\ref{auxiliary} we extend our approach to the case of an additional auxiliary system, much in the spirit of the proposal put forward by Campisi {\it et al.} in Ref.~\refcite{CampisiNJP}. Finally, Sec.~\ref{conclusions} summarises our findings and discusses the remaining open questions in this tantalising area.

\section{The interferometric scheme}
\label{description}

\begin{figure}[pt]
\centerline{{\bf (a)}\hskip5cm{\bf (b)}}
\centerline{\includegraphics[width=0.5\columnwidth]{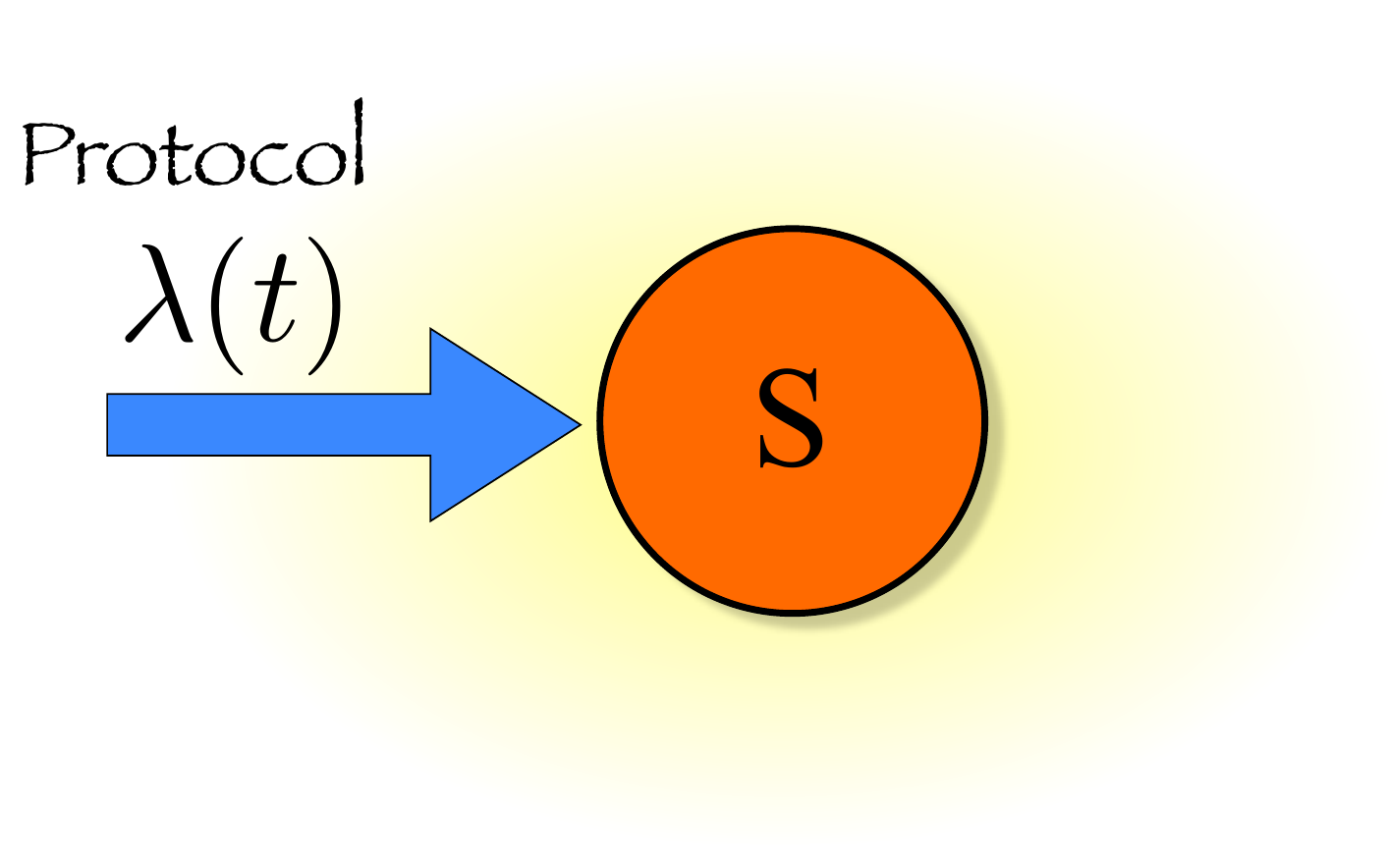}\includegraphics[width=0.5\columnwidth]{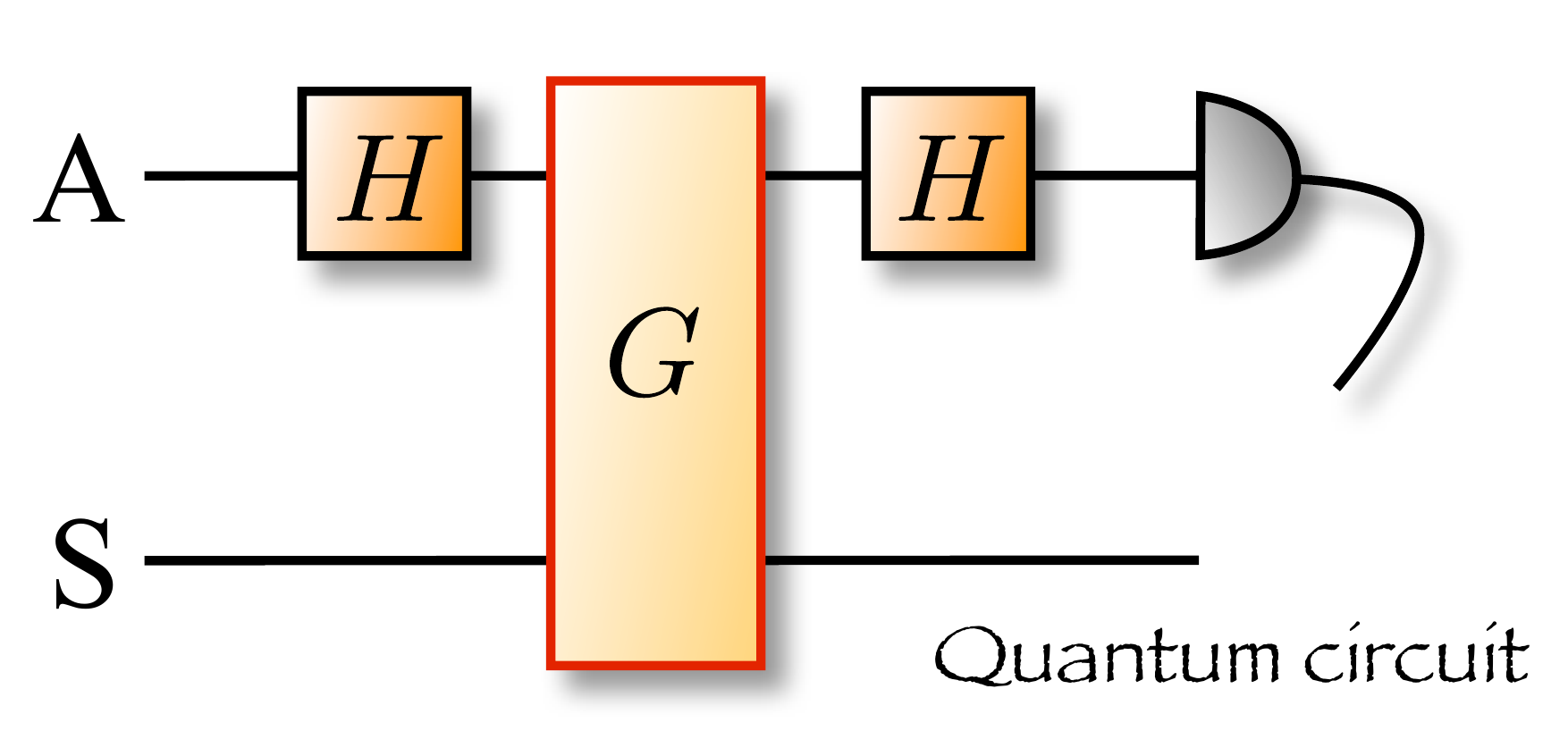}} 
\vspace*{8pt}
\caption{{\bf (a)} Pictorial sketch of a {\it protocol} embodied by the change in a parameter $\lambda(t)$ of a quantum system. {\bf (b)} Quantum circuit diagram of the interferometric approach to the reconstruction of the characteristic function. $H$ is a Hadamard gate, while $\hat G$ embodies the system-ancilla interaction whose form is specified by the specific protocol to implement. We include the symbol for the measurement of the ancilla state.}
\label{sketch}
\end{figure}

Let us consider the situation illustrated pictorially in Fig.~\ref{sketch} {\bf (a)}. A system $S$ with `bare' Hamiltonian $\hat{\cal H}_B$, describing its free evolution, is affected by a {\it protocol} described by a Hamiltonian $\hat{\cal H}_P(\lambda_t)$, which depends on an externally controlled {\it work parameter} $\lambda(t)\equiv\lambda_t$, so that the total Hamiltonian is $\hat{\cal H}_S(\lambda_t)=\hat{\cal H}_B+\hat{\cal H}_P(\lambda_t)$.
%A protocol is performed on the system, 
%For a quasistatic protocol, the first law of thermodynamics 
%states that the work done on the system is simply
%$W=H(\lambda_\tau)-H(\lambda_0)$.
We assume that at the initial time $t=0^-$ the system is in contact with a bath at inverse temperature $\beta$, so that $S$ is initialised in the thermal state 
\begin{equation}
\rho^{th}_{S}(0^-)=\frac{\e{-\beta\hat{\cal H}_S(\lambda_0)}}{{\cal Z}(\lambda_0)}.
\end{equation}
Here $\lambda_0$ is the initial value of the external parameter and $\mathcal{Z}(\lambda_t)=\textrm{Tr}{\e{-\beta\hat{\cal H}_S(\lambda_t)}}$ is the partition function. At $t=0^+$, $S$ is detached from the reservoir, while the {protocol} bringing $\lambda_t$ from $\lambda_0$ to its final value $\lambda_\tau$ starts. In order to define the probability distribution of work and its characteristic function, it is useful to write the Hamiltonian $\hat{\cal H}_S(\lambda_t)$ at the initial and final time of the protocol in terms of the corresponding spectral decomposition. That is
\begin{equation}
\hat{\cal H}_S(\lambda_0)=\sum_n E_n(\lambda_0) \ket{n}\bra{n}~~\text{and}~~\hat{\cal H}_S(\lambda_\tau)=\sum_M E'_M(\lambda_\tau) \ket{M}\bra{M},
 \end{equation}
where $E_n$ ($E'_M$) is the $n^\text{th}$ ($M^\text{th}$) eigenvalue of the initial (final) Hamiltonian associated with the eigenvector $\ket{n}$ ($\ket{M}$). The corresponding work distribution can be written as~\cite{Tasaki}
\begin{equation}
P(W)=\sum_{n,M} p(n,M) \delta\left[W-(E'_M-E_n)\right].
\label{eq:qworkdist}
\end{equation}
Here $p(n,M)=\mathrm{Tr}[\ket{M}\bra{M}\hat U_\tau \ket{n}\bra{n}\rho_S\ket{n}\bra{n}\hat U^\dagger_\tau]$ is the joint probability of finding the system in $\ket{n}$ at time $t=0$ and in state  $\ket{M}$ at time $\tau$, after the evolution ruled by the time-propagator $\hat U_\tau$. %Such probability can be decomposed in the product $p(n,m)=p^0_n\;  p^\tau_{m \vert n}$, with $p^0_n$ the probability that the system is found in state $\ket{n}$ at time $t=0$ and the conditional probability $p^\tau_{m|n}$ to find it in $\ket{m}$ at time $\tau$ if it was initially in $\ket n$.%
Obviously, such a joint probability can be decomposed as $p(n,M)=p^0_n\;  p^\tau_{M \vert n}$, where $p_n^0$ is the probability that the system is found in state $\ket{n}$ at time $t=0$ and $p^\tau_{M\vert n}$ is the conditional probability to find $S$ in $\ket{M}$ at time $\tau$ if it was initially in $\ket n$.
Therefore, $P(W)$ bears information on the statistics of the initial state and the fluctuations arising from quantum dynamics and measurement statistics. The characteristic function of the work probability distribution of $P(W)$ is then defined as\cite{lutz}
\begin{equation}
%\begin{aligned}
\chi(u)=\int\!{dW}\e{iuW}P(W)=\tr{\hat{U}^\dag_\tau\e{iu\hat{\cal H}_S(\lambda_\tau)}\hat U_\tau\e{-iu\hat{\cal H}_S(\lambda_0)}\rho^{th}_\textrm{S}(\lambda_0)}.
\label{characteristicfunction}
%\end{aligned}
\end{equation}
%where $\hat{U}_\tau$ is the unitary operator describing the system evolution.

We shall now recall the interferometric scheme for the determination of $\chi(u)$ presented in Ref.~\refcite{Mazzola}. 
Let us introduce an ancillary qubit $A$ encoded in the energy states of a two-level system $\{\ket{0}_A,\ket{1}_A\}$. We prepare the ancilla in $\ket{0}_A$ and apply a Hadamard transform $\hat H=(\hat\sigma_x+\hat\sigma_z)/\sqrt2$~\cite{Nielsen} that changes such state into $\ket{+}_A=(\ket{0}_A+\ket{1}_A)/\sqrt{2}$. We then apply the system-ancilla evolution operator  
\begin{equation}\label{Ggenerale}
\hat G(u)=\hat U_\tau e^{-i \hat{\cal H}_S^i u}\otimes\ket{0}\!\bra{0}_A+e^{-i \hat{\cal H}_S^f u}\hat U_\tau\otimes\ket{1}\!\bra{1}_A,
\end{equation}
to their joint state $\ket{+}\!\bra{+}_A\otimes\rho^{th}_S$. In Eq.~\eqref{Ggenerale} we have introduced the notation $\hat{\cal H}_S^i=\hat{\cal H}_S(\lambda_0)$ [$\hat{\cal H}_S^f=\hat{\cal H}_S(\lambda_\tau)$]. We then subject $A$ to a second Hadamard transform and trace over the degree of freedom of the system. The ancilla is correspondingly found in a state that depends on $\chi(u)$ as
\begin{equation}
\label{reduced}
\begin{aligned}
\rho_A&=\textrm{Tr}_S[\hat H\hat{G}(u)(\ket{+}_A\bra{+}\otimes\rho^{th}_S)\hat{G}^\dag(u)\hat H]=(\hat I_A+\alpha\hat\sigma_z+\nu\hat\sigma_y)/2
\end{aligned}\end{equation}
with $\alpha=\textrm{Re}\chi(u)$ and $\nu=\textrm{Im}\chi(u)$, and $\hat{\sigma}_{x,y,z}$ the Pauli operators of the ancillary qubit. From this analysis it should be clear that $\chi(u)$ can be easily reconstructed by measuring the longitudinal and transverse magnetization $\langle\hat\sigma_{z,A}\rangle$ and $\langle\hat\sigma_{y,A}\rangle$ for every value of $u$ deemed necessary.
Notice that $\hat{G}(u)$ can be decomposed into local transformations and $A$-controlled gates as $\hat{G}(u)=(\hat I_S\otimes\hat\sigma_{x,A})\hat{G}_2(u)(\hat I_S\otimes\hat\sigma_{x,A})\hat G_1(u)$ with %$\hat V(u)=\e{-i\hat{\cal H}_fu}\e{-i\hat{\cal H}_iu}\otimes{\hat\openone_A}$ and
\begin{equation}
\label{sequence}
\begin{aligned}
\hat G_{1}(u)&=\hat I_S\otimes\ket{0}\!\bra{0}_A+\e{-i\hat{\cal H}_S^fu}\hat{U}_\tau\otimes\ket{1}\!\bra{1}_A,\\
\hat G_{2}(u)&=\hat I_S\otimes\ket{0}\!\bra{0}_A+\hat{U}_\tau\e{-i\hat{\cal H}_S^iu}\otimes\ket{1}\!\bra{1}_A.
\end{aligned}
\end{equation}
Much in analogy with the inference of a relative phase originated by a dynamics in a phase-estimation protocol, our scheme clearly relies on the interference between orthogonal `evolution paths' of the system, which are to interfere at the end of the scheme (thanks to the mixing operated by the second Hadamard gate) and are imprinted in the state of the ancilla. The scheme is reminiscent of a Ramsey-like interferometer\cite{Ramsey}, which thus motivates and justify our claim for an interferometric approach to the reconstruction of the work statistics following a process. 

We emphasise that we did not make any assumptions on the form of $\hat{\cal H}_S(\lambda_t)$. In fact we allow the Hamiltonian to not commute with itself (at different instant of time) and with the unitary evolution operator. That is, our approach can be equally adopted in the cases %namely
%\begin{equation}
$[\hat{\cal H}^i_S,\hat{\cal H}_S^f]\neq 0$ and $[\hat U_\tau,\hat{\cal H}^{i(f)}_S]\neq0$.
%\end{equation}
However if such commutators are null, a much simpler version of the protocol holds, as the conditional gate in Eq.~\eqref{Ggenerale} simply becomes
\begin{equation}\label{Gsimple}
\hat G^S(u)=e^{-i \hat{\cal H}_S^i u}\otimes\ket{0}\!\bra{0}_A+e^{-i \hat{\cal H}_S^f u}\otimes\ket{1}\!\bra{1}_A.
\end{equation}
The protocol proceeds exactly as described above with the the replacement $\hat{G}(u)\rightarrow\hat{G}^S(u)$. As shown above, $\hat{G}^S(u)$ can be split in two $A$-controlled gates and local transformation (as in the general case) as $\hat{G}^S(u)=(\hat I_S\otimes\hat\sigma_{x,A})\hat{G}^S_2(u)(\hat I_S\otimes\hat\sigma_{x,A})\hat G^S_1(u)$ with 
\begin{equation}
\label{Gsimpleparts}
\begin{aligned}
\hat G^S_{1}(u)&=\hat{I}_S\otimes\ket{0}\!\bra{0}_A+\e{-i\hat{\cal H}_S^fu}\otimes\ket{1}\!\bra{1}_A,\\
\hat G^S_{2}(u)&=\hat{I}_S\otimes\ket{0}\!\bra{0}_A+\e{-i\hat{\cal H}_S^iu}\otimes\ket{1}\!\bra{1}_A.
\end{aligned}
\end{equation}

%Our approach is to give a set of gates that implement the protocol, these gates are not unique, actually as we demonstrate later slightly different gates can implement the protocol as well. Nevertheless in various cases it is possible to demonstrate that the joint evolution of a system and an ancilla suitably coupled generate naturally the desire evolution.

\section{Physical example}
\label{Example}

\begin{figure}[pt]
\centerline{{\bf (a)}\hskip5cm{\bf (b)}}
\centerline{\includegraphics[width=1.1\columnwidth]{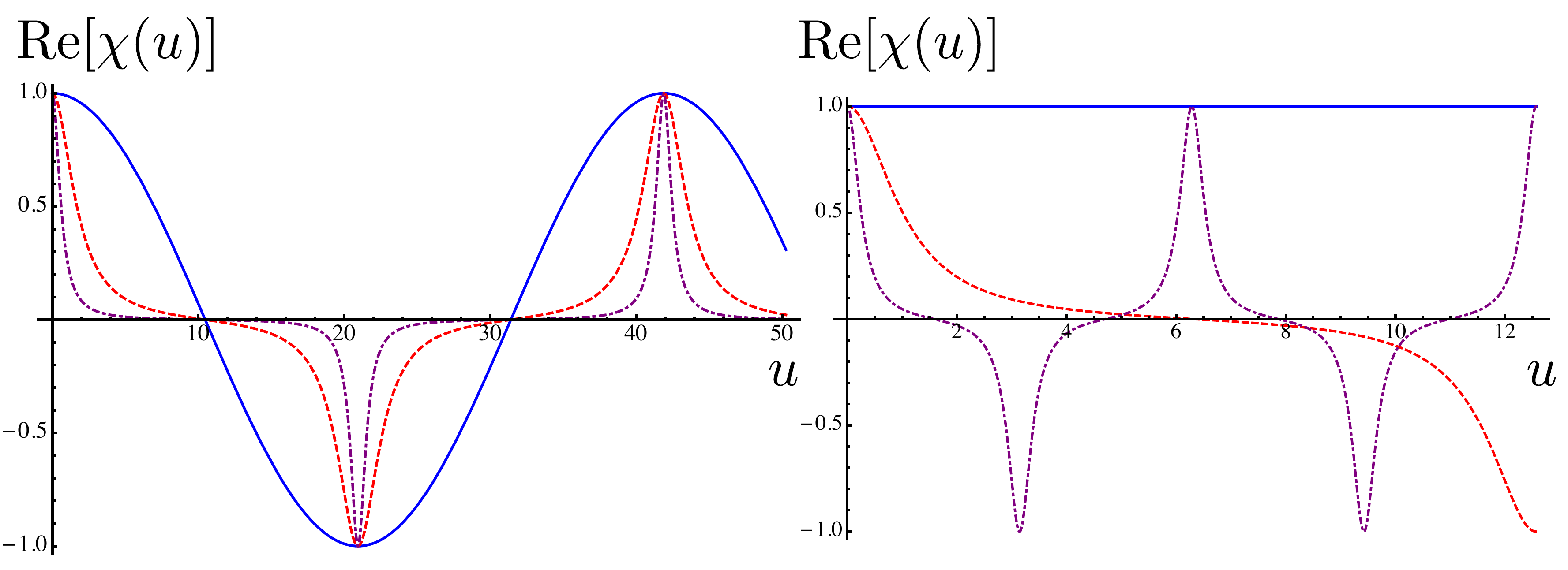}}
\vspace*{8pt}
\caption{Real part of the characteristic function of work distribution for the process described in the body of the paper. Panel {\bf (a)} We study the cases of $\Delta\lambda=0.3$ with $\nbar=0$ (solid blue line), $\nbar=1.5$ (dashed red one), and $\nbar=5$ (dot-dashed purple line). {\bf (b)} We complement our study by looking at the case of $\nbar=1.5$ with $\Delta\lambda=0$ (solid blue line), $\Delta\lambda=0.5$ (dashed red line), and $\Delta\lambda=2$ (dot-dashed purple one).}
\label{characteristicRe}
\end{figure}

\begin{figure}[pt]
\centerline{{\bf (a)}\hskip5cm{\bf (b)}}
\centerline{\includegraphics[width=1.1\columnwidth]{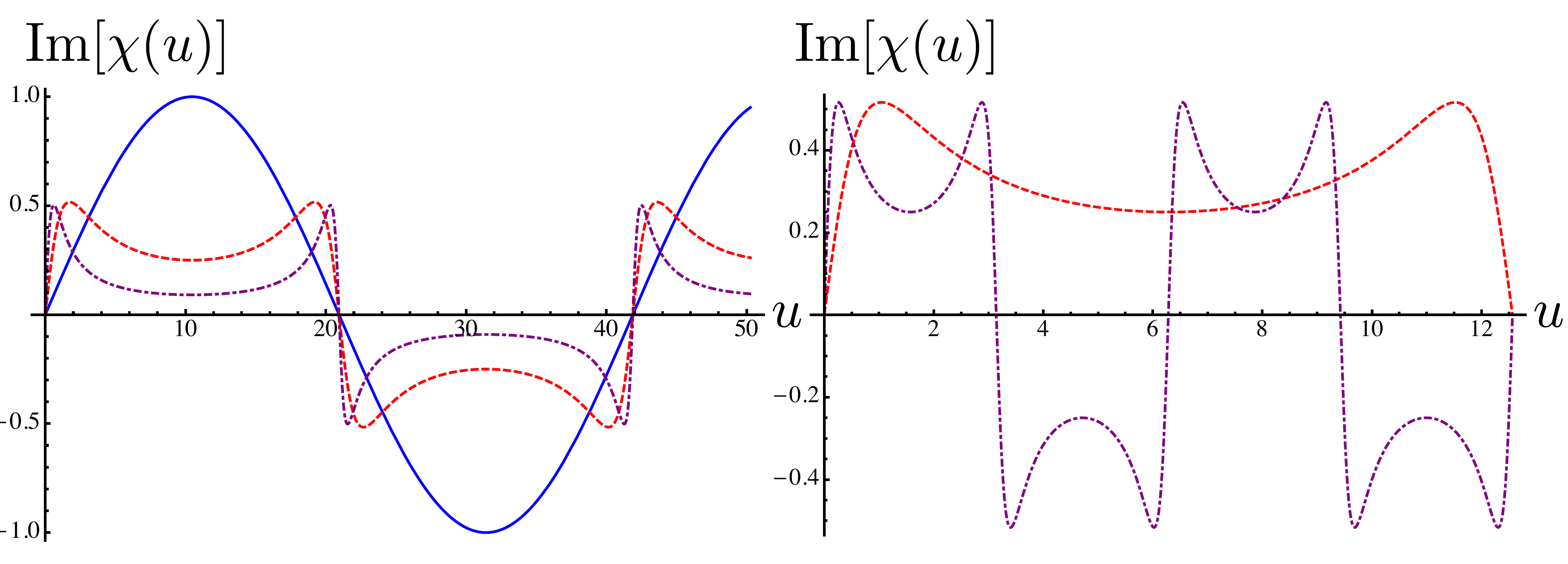}}
\vspace*{8pt}
\caption{Imaginary part of the characteristic function of work distribution for the process described in the body of the paper. Panel {\bf (a)} We study the cases of $\Delta\lambda=0.3$ with $\nbar=0$ (solid blue line), $\nbar=1.5$ (dashed red one), and $\nbar=5$ (dot-dashed purple line). {\bf (b)} We complement our study by looking at the case of $\nbar=1.5$ with $\Delta\lambda=0$ (solid blue line), $\Delta\lambda=0.5$ (dashed red line), and $\Delta\lambda=2$ (dot-dashed purple one).}
\label{characteristicIm}
\end{figure}

We now discuss an explicit physical scenario in which our interferometric scheme can be applied and illustrated efficiently. In details, we consider a harmonic oscillator whose frequency, representing the work parameter $\lambda_t$ of the protocol, is changed in time according to a chosen functional form. The Hamiltonian of the system is thus
\begin{equation}
\hat H_S=\hbar\lambda_t(\hat a^\dag\hat a+1/2)
\end{equation}
with $\lambda_t$ that is changed in time. For illustrative purposes, here we concentrate on the case of a sudden quench of the frequency of the oscillator\cite{CampisiJPhysA}, the case of a generic temporal dependence being only a generalisation of the forthcoming discussion. It is straightforward to check that, in this case, the characteristic function of the work distribution following the frequency quench is given by 
\begin{equation}
\label{chi}
\chi(u)=\sum^\infty_{n=0}\frac{\nbar^n}{(1+\nbar)^{n+1}}e^{iu \Delta\lambda (n+1/2)}=\frac{e^{\frac{iu \Delta\lambda}{2}}}{1+\nbar(1-e^{iu \Delta\lambda })},
\end{equation}
where $\Delta\lambda=\lambda_\tau-\lambda_0$ and $\nbar$ is the mean number of excitations in the initial thermal state $\rho^{th}_S(0^-)=\sum^\infty_{n=0}\frac{\nbar^n}{(1+\nbar)^{n+1}}\ket{n}\bra{n}$. This function depends crucially on the amplitude of the quench $\Delta\lambda$ and the mean thermal number $\nbar$. For no quench (i.e. $\Delta\lambda=0$), $\chi(u)=1$ and no work is done on the harmonic oscillator. On the other hand, there are values of the quench amplitude that correspond to the occurrence of `resonances' in $\chi(u)$, as it is seen from Figs.~\ref{characteristicRe} and \ref{characteristicIm}. The consequences of such dependences are more clearly seen from the form taken by the corresponding $P(W)$ and the average work $\langle W\rangle$. 

\begin{figure}[pt]
\centerline{{\bf (a)}\hskip5cm{\bf (b)}}
\centerline{\includegraphics[width=0.55\columnwidth]{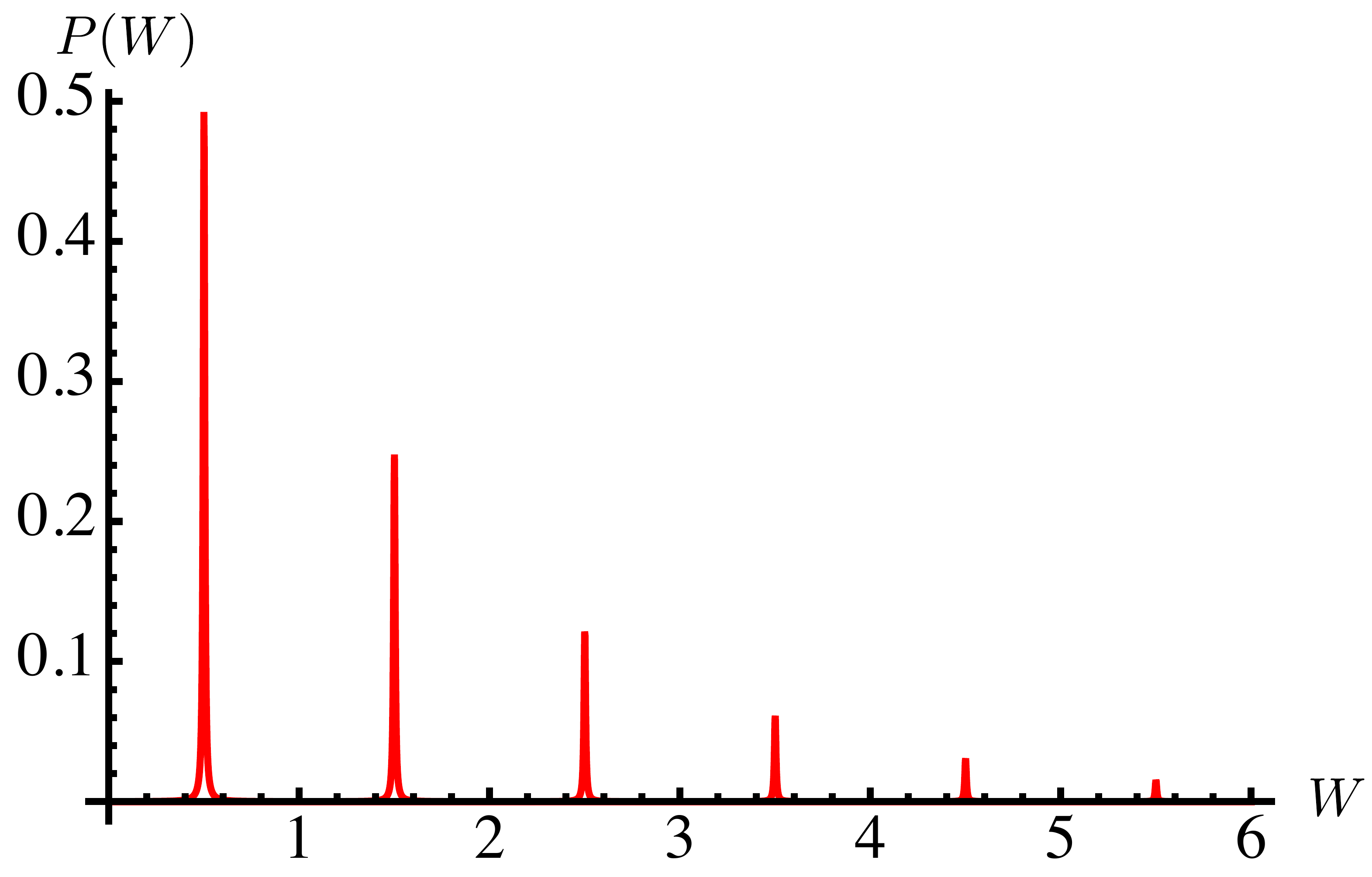}\includegraphics[width=0.55\columnwidth]{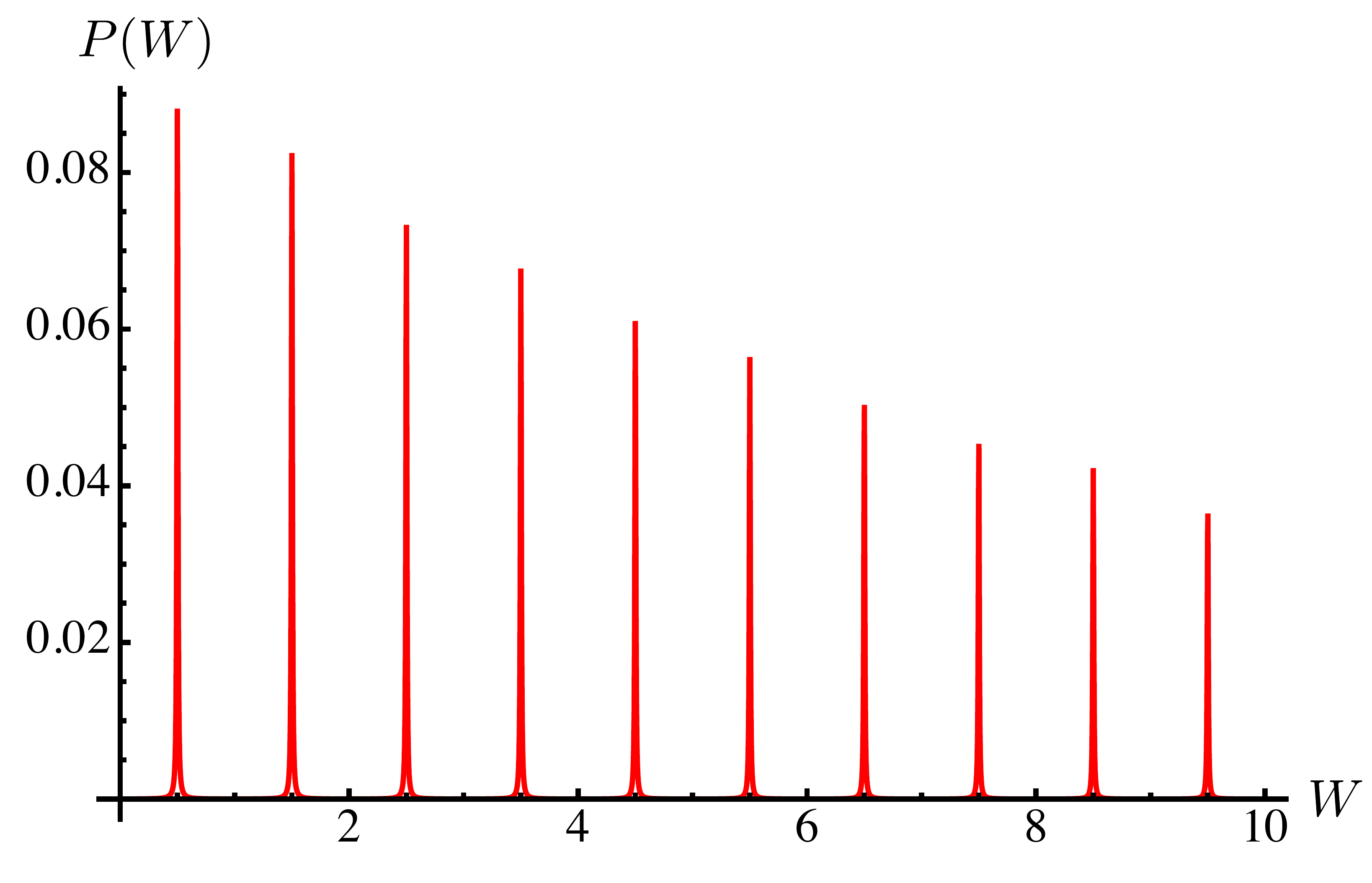}}
\vspace*{8pt}
\caption{Work probability distribution for $\Delta\lambda=1$ and $\nbar=1$ [panel {\bf (a)}] and $\nbar=10$ [panel {\bf (b)}]. For easiness of illustration, the expected Dirac delta functions are here replaced by very narrow Lorenzian functions, centred at $(n+1/2)\Delta\lambda$.}
\label{distri}
\end{figure}

The first is determined by taking the anti-Fourier transform of $\chi(u)$, in line with the definition given in Eq.~\eqref{characteristicfunction}. In order to gather analytic insight into this problem, we have evaluated the integral
\begin{equation}
{\cal I}(\epsilon)=\frac{1}{2\pi}\int^\epsilon_{-\epsilon}\chi(u)e^{-iWu}du,
\end{equation}
which is such that $P(W)=\lim_{\epsilon\to\infty}{\cal I}(\epsilon)$. We get
\begin{equation}
{\cal I}(\epsilon)=-\frac{ie^{-iW\epsilon}}{\pi(1+\nbar)(2W-\Delta\lambda)}\left[e^{\frac{i}{2} \epsilon(4W-\Delta\lambda)}{}_{2}F_{1}(1,a;1+a,z)-e^{\frac{i}{2}\epsilon\Delta\lambda}{}_{2}F_{1}(1,a;1+a,z^*)\right]
\end{equation}
with $a=1/2-W/\Delta\lambda$, $z=e^{-i\epsilon\Delta\lambda}\nbar/(1+\nbar)$, and ${}_{2}F_1(\alpha,\beta;\gamma,z)$ the Hypergeometric function. For $\nbar=0$, we have $z=0$ and ${}_2F_1(\alpha,\beta;\gamma,0)=1$. This implies that 
\begin{equation}
{\cal I}(\epsilon)=-\frac{i}{\pi(2W-\Delta\lambda)}\left[e^{i\epsilon(W-\Delta\lambda/2)}-e^{-i\epsilon(W-\Delta\lambda/2)}\right]=\frac{2\sin[\epsilon(W-\Delta\lambda/2)]}{\pi(W-\Delta\lambda)}.
\end{equation}
Therefore, $\lim_{\epsilon\to\infty}{\cal I}(\epsilon)=\delta(W-\Delta\lambda/2)$, which is in line with the physical expectations at null temperature. For $\nbar\neq0$, on the other hand, it is convenient to use the power-series definition of the Hypergeometric function, i.e. ${}_{2}F_1(\alpha,\beta;\gamma,z)=\sum^\infty_{n=0}\frac{(\alpha)_n(\beta)_n}{(\gamma)_n}\frac{z^n}{n!}$ with $(q)_n$ the Pochhammer symbol of argument $q$. For the specific case at hand, we have that
\begin{equation}
{}_{2}F_1(1,a;1+a,z)=\sum^\infty_{n=0}\frac{a}{a+n}z^n=\sum^\infty_{n=0}\frac{a}{a+n}\frac{\nbar^n}{(1+\nbar)^{n+1}}e^{-in\epsilon\Delta\lambda}.
\end{equation}
In turn, this implies that the probability distribution of work is made out of Dirac-delta peaks centred at $(n+1/2)\Delta\lambda$ and of amplitude $\nbar^n/(1+\nbar)^{n+1}$, which are dictated by the statistics of the initial thermal state. Two instances of such distribution are illustrated in Fig.~\ref{distri}, where we see that higher temperatures correspond to the emergence of many peaks in $P(W)$ due to the large number of states entering $\rho^{th}_S(0^-)$ and a correspondingly large number of state transitions.

As for the average work, this can be easily determined using the characteristic function $\chi(u)$ as $\langle W\rangle=-\left.i\partial_u\chi(u)\right|_{u=0}$, whose explicit evaluation gives us
\begin{equation}
\langle W\rangle=\Delta\lambda\left(\nbar+\frac12\right),
\end{equation}
showing that the average work linearly increases with the amplitude of the quench and the mean thermal number of excitations.

%Our scheme requires the introduction of an ancillary qubits, which plays the role of control in the conditional gates of Eq. \eqref{Ggenerale}. 

We now show that a suitable coupling between the system and an ancilla qubit allows us to generate the conditional operations, introduced in the previous Section, necessary for the interferometric reconstruction of the characteristic function. In order to fix the ideas, we can think of the harmonic oscillator as embodied by the fundamental flexural mode of a suspended double-clamped cantilever. %Such mechanical mode undergoes a protocol that changes its frequency in time. 
The ancillary qubit needed to apply our scheme can be provided by a Cooper-pair box capacitively coupled to the oscillator. The Hamiltonian for such superconducting-mechanical system reads\cite{Armour}
\begin{equation}\label{exampleini}
\hat{\tilde{H}}_{SA}=\frac{\hbar\epsilon_0}{2}\hat{\sigma}_z+\hbar\delta\hat{\sigma}_x+\hbar\omega\left(\hat{a}^\dag\hat{a}+\frac{1}{2}\right)+\hbar g(\hat{a}+\hat{a}^\dag)\otimes\hat{\sigma}_z
\end{equation}
where $\epsilon_0$ and $\delta$ are the qubit energy scales, $\omega$ is the resonator frequency, $g$ is the coupling constant between resonator and qubit, and any spin operator refers to the ancilla. %while $\hat{a}$ ($\hat{a}^\dag$) refers to the oscillator.
The following working conditions allow to derive a simplified Hamiltonian model: we assume a wide separation of time scales for the mechanical oscillator and the superconducting qubit, i.e. $\delta\gg\omega$. Moreover, we assume that the coupling term is a weak perturbation with respect to the free evolution, so that the rotating wave approximation can be invoked. %, namely $g^2/2\Delta\ll\hbar\omega$
Finally, we assume that $\epsilon_0$ is tuned to be null (this can be done by replacing the Cooper-pair box with a superconducting quantum interference device (SQUID), pierced by an external magnetic field). In this regime, the resulting effective dispersive Hamiltonian reads
\begin{equation}\label{example}
\hat{H}_{SA}(t)=\hbar\omega\left(\hat{a}^\dag \hat{a}+\frac{1}{2}\right)+\hbar\delta \hat{\sigma}_z+\hbar\lambda_t\left(\hat{a}^\dag \hat{a}+\frac{1}{2}\right)\otimes\hat{\sigma}_z,
\end{equation}
where $\lambda_t=g^2/\delta$ is the work parameter that is changed during the protocol. %Notice that $\hat{\sigma}_{x,y,z}$ are now written in a different basis compared to the Hamiltonian in Eq. \eqref{exampleini}. 
We can write Eq.~\eqref{example} in a compact way as $\hat{H}_{SA}(t)=\hat{H}_S+\hat{H}_A+\hat{H}_S(\lambda_t)\otimes\hat{\sigma}_z$ with obvious meaning of each term.
Such model commutes with itself at every instant of time and with the unitary evolution operator, so that the stream-lined version of the gate-decomposition given in Eq.~\eqref{Gsimple} can be used. We can split the time-evolution operators in three different parts: The first two are just the free evolutions of system and ancilla, while the third one describes the effects of the interaction between the two. Explicitly, we have %
%We notice that the joint evolution of system and ancilla caused by such Hamiltonian for a generic time-dependence of the work parameter $\lambda_t$ is given by
\begin{equation}
\label{USA}
\hat U_{SA}(t)=\hat{\mathcal{T}}e^{-i \int_0^\tau \hat H_{SA}(t)dt}=e^{-i \hat H_S\tau}e^{-i \hat H_A\tau}e^{-i\int_0^\tau \hat H_S(\lambda_t)dt\otimes\hat \sigma_z},
\end{equation}
which can be expanded in power series as
\begin{equation}
\begin{aligned}
e^{-i\int_0^\tau\hat H_S(\lambda_t) dt \otimes\hat \sigma_z}%&=\sum_{n=0}^\infty \frac{\left(-i\int_0^\tau \hat H_S(\lambda_t) dt\right)^n}{n!}\otimes\hat \sigma_z^n\\
&=\sum_{n=0}^\infty \frac{\left(-i\int_0^\tau \hat H_S(\lambda_t) dt\right)^{2n}}{(2n)!}\otimes\hat I_A+\sum_{n=0}^\infty \frac{\left(-i\int_0^\tau \hat H_S(\lambda_t) dt\right)^{2n+1}}{(2n+1)!}\otimes\hat \sigma_z\\
&=\cos\left(\int_0^\tau \hat H_S(\lambda_t)dt\right)\otimes\hat I_A-i \sin\left(\int_0^\tau \hat H_S(\lambda_t)dt\right)\otimes\hat \sigma_z.
\end{aligned}
\end{equation}
%where we used $\hat \sigma_z^{2n+1}=\hat \sigma_z$ and $\hat \sigma_z^{2n}=\hat I$.
By plugging this expression in Eq. \eqref{USA} and using the Euler's formula, $\hat U_{SA}(t)$ can be rewritten as
\begin{equation}
\hat U_{SA}(t)=\left(e^{-i\int_0^\tau \hat H_S(\lambda_t)dt}\ket{0}\bra{0}_A+e^{i\int_0^\tau \hat H_S(\lambda_t)dt}\ket{1}\bra{1}_A\right)e^{-i(\hat H_S+\hat H_A)\tau}.
\end{equation}
This expression has much in common with the one in Eq. \eqref{Gsimple}. We notice that the terms $\mathcal{\hat{H}}_{S}^{i,f}$ in Eqs.~\eqref{Ggenerale}-\eqref{Gsimpleparts} correspond here to $\mathcal{\hat{H}}_{S}^{i(f)}=\hat{H}_S+\hat H_S\left(\lambda_{0(\tau)}\right)$.
Consider the following two gates 
\begin{equation}\begin{split}
\mathcal{\hat G}_1(u)=&e^{-i  \hat H_{SA}(\tau)u/2}=\left[e^{-i  \hat H_S(\lambda_\tau)u/2}\ket{0}\bra{0}_A+e^{i  \hat H_S(\lambda_\tau)u/2}\ket{1}\bra{1}_A\right]e^{-i(\hat H_S+\hat H_A)u/2},\\
\mathcal{\hat G}_2(u)=&e^{-i \hat  H_{SA}(0)u/2}=\left[e^{-i  \hat H_S(\lambda_0)u/2}\ket{0}\bra{0}_A+e^{i  \hat H_S(\lambda_0)u/2}\ket{1}\bra{1}_A\right]e^{-i(\hat H_S+\hat H_A)u/2}.
\end{split}\end{equation}
These are the result of a joint evolution of system and ancilla for a time $u/2$ fixing the work parameter $\lambda_t$ at its initial and its final value, respectively.
Combining these two gates as prescribed above, we obtain 
\begin{equation}\begin{aligned}
\mathcal{\hat{G}}(u)&=(\hat I_S\otimes\hat\sigma_{x,A})\mathcal{\hat{G}}_2(u)(\hat I_S\otimes\hat\sigma_{x,A})\mathcal{\hat G}_1(u)\\
&=\left[e^{-i  \left(\hat H_S(\lambda_\tau)-\hat H_S(\lambda_0)\right)\frac{u}{2}}\ket{0}\bra{0}_A+e^{i  \left(\hat H_S(\lambda_\tau)-\hat H_S(\lambda_0)\right)\frac{u}{2}}\ket{1}\bra{1}_A\right]e^{-i(\hat H_S+\hat H_A)u}.
\end{aligned}\end{equation}
This gate differs from the one in Eq. \eqref{Gsimple} for the application of local unitaries, namely
\begin{equation}
\hat G^S(u)=\hat \sigma_x\mathcal{\hat{G}}(u)\hat \sigma_x e^{i \hat H_A u}e^{-i  \left(\hat H_S(\lambda_0)+\hat H_S(\lambda_\tau)\right)u/2}.
\end{equation}
It is rather straightforward to show that the protocol produces exactly the same results of Eq. \eqref{reduced} if instead of applying $\hat G^S(u)$ we use $\mathcal{\hat{G}}(u)$.

This example demonstrates the existence of a wider class of gates than the one given in Eqs.~\eqref{Ggenerale}-\eqref{Gsimpleparts}. In fact, any gate that differs from Eq.~\eqref{Ggenerale} for local unitaries on the system and the ancilla as in
\begin{equation}\begin{aligned}
\hat G^G(u)&=\hat G(u)\mathcal{K}_S(u)\mathcal{L}_A(u)\\
&=\left(\hat U_\tau e^{-i \hat{\cal H}_S^i u}\otimes\ket{0}\!\bra{0}_A+e^{-i \hat{\cal H}_S^f u}\hat U_\tau\otimes\ket{1}\!\bra{1}_A\right)\mathcal{K}_S(u)\mathcal{L}_A(u),
\end{aligned}\end{equation}
can be equivalently used for the realisation of our scheme. In fact, when we apply this gate to $\ket{+}\!\bra{+}_A\otimes\rho^{th}_S$, the local unitary operations would cancel out. 

We are now in a position to demonstrate the effectiveness of our scheme. In fact, for a sudden quench in the frequency of the oscillator as the one addressed earlier in this Section, a rather straightforward calculation leads to the following state of the ancilla at the end of the protocol
\begin{equation}
\begin{aligned}
\rho_A&=\frac12
\begin{pmatrix}
1+\sum^\infty_{n=0}\dfrac{\nbar^n\cos[u\Delta\lambda(n+1/2)]}{(1+\nbar)^{n+1}}&-i\sum^\infty_{n=0}\dfrac{\nbar^n\sin[u\Delta\lambda(n+1/2)]}{(1+\nbar)^{n+1}}\\
i\sum^\infty_{n=0}\dfrac{\nbar^n\sin[u\Delta\lambda(n+1/2)]}{(1+\nbar)^{n+1}}&1-\sum^\infty_{n=0}\dfrac{\nbar^n\cos[u\Delta\lambda(n+1/2)]}{(1+\nbar)^{n+1}}
\end{pmatrix}\\
&=\frac12
\begin{pmatrix}
1+\dfrac{\cos[u\Delta\lambda/2]}{1+2\nbar(1+\nbar)(1-\cos[u\Delta\lambda])}&-i\dfrac{(1+2\nbar)\sin[u\Delta\lambda/2]}{1+2\nbar(1+\nbar)(1-\cos[u\Delta\lambda])}\\
i\dfrac{(1+2\nbar)\sin[u\Delta\lambda/2]}{1+2\nbar(1+\nbar)(1-\cos[u\Delta\lambda])}&1-\dfrac{\cos[u\Delta\lambda/2]}{1+2\nbar(1+\nbar)(1-\cos[u\Delta\lambda])}
\end{pmatrix},
\end{aligned}
\end{equation}
in line with the general form in Eq.~\eqref{reduced}. It is then easy to check, using Eq.~\eqref{chi}, that 
\begin{equation}
\text{Re}[\chi(u)]=\frac{\cos[u\Delta\lambda/2]}{1+2\nbar(1+\nbar)(1-\cos[u\Delta\lambda])},~~\text{Im}[\chi(u)]=\frac{(1+2\nbar)\sin[u\Delta\lambda/2]}{1+2\nbar(1+\nbar)(1-\cos[u\Delta\lambda])}.
\end{equation}
%We noticed already something similar in the example provided in Ref. \cite{Mazzola}. 
%This demonstrates that the gates that we have used here belong to a wider classes of schemes whose most general expression we do not know, yet. However this suggest the versatility of our concept and the possibility to apply it with variation to infer other quantities.

\section{Introducing an auxiliary system}
\label{auxiliary}

In Ref.~\refcite{Zanardi} a generalisation of the concept of probability distribution of work to the open case has been proposed. We go through the derivation of this expression here and show how our interferometric scheme can be applied also to infer such a generalised quantity. We believe however that this quantity represents the generalisation of the probability distribution of quantum work only in the weak interaction case. The definition of work and heat in the open quantum scenario is a difficult problem that goes beyond the scope of this work. Here we just want to demonstrate how our protocol can be adapted to detect other quantities of interest.

Consider a quantum system $S$ that, at time $t=0$, is affected by an external driving and an auxiliary system $E$ (which could be a surrounding environment) for a time $\tau$. We assume the total Hamiltonian to be written as $\hat{\mathcal{H}}_{tot}(t)=\hat{\mathcal{H}}_S(\lambda_t)+\hat{\mathcal{H}}_{SE}$, where $\hat{\mathcal{H}}_S(\lambda_t)$ contains the bare system Hamiltonian and the Hamiltonian describing the protocol, and $\hat{\mathcal{H}}_{SE}$ describes the interaction between $S$ and $E$ and the bare Hamiltonian of $E$. The {\it gedankenexperiment} proceeds as usual. We make energy measurements on $S$ at the initial and final time. The probability distribution of work is then constructed from the difference of final and initial energy weighted by the probabilities that such energy jumps occur, analogously to Eq. \eqref{eq:qworkdist}. That is
\begin{equation}\label{probdistrWE}
P_S(W)=\sum_{n,M} p_S(n,M) \delta\left[W-(E_M'-E_n)\right].
\end{equation}
However, here the joint probability is calculated taking into account the degrees of freedom of the environment too as
\begin{equation}
%\begin{aligned}
\label{pnmE}
p_S(n,M)=\mathrm{Tr}_{SE}\Big[\ket{M}\bra{M}_S\hat U_{SE}(\tau,0)\ket{n}\bra{n}_S\hat I_E(\rho_S\otimes\rho_E)\ket{n}\bra{n}_S\hat U^\dagger_{SE}(\tau,0)\Big],
%\end{aligned}
\end{equation}
where $\hat U_{SE}(\tau,0)$ is the evolution operator associated to $\hat{\cal H}_{tot}$. The rest of the notation is strictly analogous to the one used in the previous Sections. %and $(E_n,\ket{n})$ and $(E'_m,\ket{m})$ are as above the eigenvalue-eigenstate pairs of the initial and final Hamiltonian of the system only, $\hat{\mathcal{H}}_S(\lambda_t)$.
By using Eqs. \eqref{pnmE} we can calculate the characteristic function of the probability distribution $P_S(W)$ in Eq. \eqref{probdistrWE}. We define $P_n=\ket{n}\bra{n}_S$ and $Q_M=\ket{M}\bra{M}_S$ and omit the identity operator $\hat I_E$ for the sake of brevity. The characteristic function then reads
\begin{equation}\begin{aligned}\label{chiE}
\chi_S(u)&=\int\!{dW}\e{iuW}P_S(W)=\sum_{n,M} p(n,M) e^{i u(E_M'-E_n)}\\
&=\mathrm{Tr}_{SE}\left[\sum_M Q_M e^{i u E_M'} \hat U_{SE}(\tau,0)
\rho_S\otimes\rho_E \sum_n P_n e^{-i u E_M} \hat U^\dagger_{SE}(\tau,0)\right]\\
&=\mathrm{Tr}_S\left[e^{i u \mathcal{\hat H}_S(\lambda_\tau)}\sum_{j,l}\bra{j}\hat U_{SE}(\tau,0)\ket{l}p_l\rho_S e^{-i u \mathcal{\hat H}_S(\lambda_0)}\bra{l}\hat U^\dagger_{SE}(\tau,0)\ket{j}\right]\\
&=\mathrm{Tr}_S\left[e^{i u \mathcal{\hat H}_S(\lambda_\tau)}\sum_{j}\hat K_j\rho_S e^{-i u \mathcal{\hat H}_S(\lambda_0)}\hat K^\dagger_j\right],
\end{aligned}\end{equation}
where where we defined $\hat K_j=\sum_l \sqrt{p_l} \bra{j}\hat U_{SE}(\tau,0)\ket{l}$. To derive this expression we assumed that the state of the system $\rho_S$ is diagonal in the basis of the initial Hamiltonian, as for a thermal state. In the third row we wrote explicitly the state of the environment in its diagonal basis as $\rho_E=\sum_l p_l\ket{l}\bra{l}_E$ and performed the trace over the environment in the same basis. 

Our interferometric scheme can be applied to detect the quantity in Eq. \eqref{chiE}. As usual we need to introduce an ancillary qubit $A$ that works as a controller, which we initialise in $\ket{+}\!\bra{+}_A$ through a Hadamard gate. At this point we apply the following gate
\begin{equation}
\hat G^{SE}=\hat U_{SE}(\tau,0) e^{-i u \mathcal{\hat H}_S(\lambda_0)}\ket{0}\bra{0}_A+e^{-i u \mathcal{\hat H}_S(\lambda_\tau)}\hat U_{SE}(\tau,0)\ket{1}\bra{1}_A
\end{equation}
to the initial state $\ket{+}\bra{+}_A\otimes\rho_S\otimes\rho_E$, after which we perform an additional Hadamard operation on the qubit. By detecting the qubit only, i.e. tracing over the degrees of the system and environment, we find that the function $\chi_S(u)$ was mapped onto the state of the ancilla exactly as in Eq.~\eqref{reduced}
\begin{equation}
\label{reduced1}
%\begin{aligned}
\rho_A=\textrm{Tr}_{SE}[\hat H\hat{G}(u)(\ket{+}\bra{+}_A\otimes\rho_S\otimes\rho_E)\hat{G}^\dag(u)\hat H]=(\hat I_A+\alpha\hat\sigma_z+\nu\hat\sigma_y)/2
%\end{aligned}
\end{equation}
with $\alpha=\textrm{Re}\chi_{s}(u)$ and $\nu=\textrm{Im}\chi_{s}(u)$. We believe that this approach can be useful to address the out-of-equilibrium statistics of a quantum open system, a problem that is currently under study and that will be the focus of forthcoming work.

\section{Conclusions and outlook}
\label{conclusions}

We have addressed the working principles and flexibility features of an interferometric protocol for the reconstruction of the work statistics of a quantum system subjected to a time-dependent process. Our proposal has already shown its handiness in the characterisation of quantum fluctuation theorems in controlled experimental situations\cite{Batalhao}, and in order to illustrate its features we have addressed a physically motivated example consisting of a quantum harmonic oscillator with variable frequency. Finally, we have briefly sketched the approach that should be used in order to reconstruct the characteristic function of work distribution for a system weakly coupled to an environment. This leaves room for further interesting questions, such as the design of experimentally viable schemes for the inference of heat exchanged in a quantum process, which are yet to be answered and are the current focus of ongoing investigations.

\section*{Acknowledgments}

We thank T. J. G. Apollaro, M. Campisi, R. Dorner, J. Goold, I. Lesanovski, K. Modi, F. Plastina, R. M. Serra, F. Semi${\rm\tilde{a}}$o, D. Soares-Pinto, and V. Vedral for invaluable discussions on the subject of this paper. LM is supported by the EU through a Marie Curie IEF Fellowship. MP thanks the Alexander von Humboldt Foundation and the UK EPSRC for a Career Acceleration Fellowship and a grant awarded under the ``New Directions for Research Leaders" initiative (EP/G004579/1). GDC and MP acknowledge the John Templeton Foundation (grant ID 43467) and the EU Collaborative Project TherMiQ (Grant Agreement 618074) for financial support.

\end{document}